\documentclass[conference]{IEEEtran}
\IEEEoverridecommandlockouts
\usepackage{cite}
\usepackage{amsmath,amssymb,amsfonts}
\usepackage{algorithmic}
\usepackage{graphicx}
\usepackage{textcomp}
\usepackage{xcolor}
\usepackage{pbalance}
\usepackage[hyphens]{url}
\usepackage{soul}
\usepackage{hyperref}
\hypersetup{
  colorlinks=true,
  linkcolor=blue!60!black,
  citecolor=purple,
  urlcolor=blue!60!black,
}
\usepackage{orcidlink}
\usepackage{xspace}
\usepackage[capitalise,noabbrev]{cleveref}
\usepackage{tikz} 
\usepackage{fancybox}
\usepackage{fancyhdr} 
\usepackage{comment}
\usepackage{enumitem}
\usepackage{multirow}
\usepackage{booktabs}
\def\BibTeX{{\rm B\kern-.05em{\sc i\kern-.025em b}\kern-.08em
    T\kern-.1667em\lower.7ex\hbox{E}\kern-.125emX}}

\begin{document}
\bstctlcite{IEEEexample:BSTcontrol}

% Ensure letter paper
\pdfpagewidth=8.5in
\pdfpageheight=11in

\author{\IEEEauthorblockN{Jeonghyun Woo\IEEEauthorrefmark{1}\,\orcidlink{0000-0001-5819-0693}}
\IEEEauthorblockA{
University of British Columbia\\
jhwoo36@ece.ubc.ca}
\and
\IEEEauthorblockN{Junsu Kim\IEEEauthorrefmark{1}\,\orcidlink{0009-0008-1582-5212}}
\IEEEauthorblockA{
University of British Columbia\\
junsukim@ece.ubc.ca}
\and
\IEEEauthorblockN{Aamer Jaleel\,\orcidlink{0000-0002-5709-2992}}
\IEEEauthorblockA{
NVIDIA\\
ajaleel@nvidia.com}
\and
\IEEEauthorblockN{Prashant J. Nair\,\orcidlink{0000-0002-1732-4314}}
\IEEEauthorblockA{
University of British Columbia\\
prashantnair@ece.ubc.ca}
\thanks{\IEEEauthorrefmark{1}Both authors contributed equally to this work.}
}
%%%%%%%%%%%---SETME-----%%%%%%%%%%%%%
\title{Loaded Dice: Solving the Non-Selection Problem for Scalable Probabilistic RowHammer Defense}
%%%%%%%%%%%%%%%%%%%%%%%%%%%%%%%%%%%%

\pagenumbering{arabic}

\maketitle

\thispagestyle{plain}
\pagestyle{plain}
%%%%%%% -- HELPER DEFINES STARTS -- %%%%%%%%%%%
\newcommand{\NRH}{$\text{N}_{\text{RH}}$\xspace}
\newcommand{\NTARD}{$\text{T}_{\text{TARD}}$\xspace}
\newcommand{\TREFI}{\text{tREFI}}
\newcommand{\TRFC}{\text{tRFC}}
\newcommand{\TREFW}{\text{tREFW}}
\newcommand{\TREF}{$\text{TREF}$\xspace}
\newcommand{\NBO}{$\text{N}_{\text{BO}}$\xspace}
\newcommand{\ABOACT}{$\text{ABO}_{\text{ACT}}$\xspace}
\newcommand{\ABODELAY}{$\text{ABO}_{\text{Delay}}$\xspace}
\newcommand{\NMIT}{$\text{N}_{\text{mit}}$\xspace}
\newcommand{\RFMAB}{$\text{RFM}_{\text{ab}}$\xspace}
\newcommand{\RFMPB}{$\text{RFM}_{\text{pb}}$\xspace}
\newcommand{\RFMSB}{$\text{RFM}_{\text{sb}}$\xspace}
\newcommand{\ALERT}{\text{Alert}}
\newcommand{\BR}{\text{BR}}
\newcommand{\RFM}{\text{RFM}}
\newcommand{\ACT}{\text{ACT}}
\newcommand{\PRE}{\text{PRE}}
\newcommand{\TRC}{\text{tRC}}
\newcommand{\TRP}{\text{tRP}}
\newcommand{\REF}{\text{REF}}

\newcommand{\NRHS}{$\text{N}_{\text{RH-S}}$\xspace}

\newcommand{\TRHS}{$\text{T}_{\text{RH-S}}$\xspace}
\newcommand{\TRHD}{$\text{T}_{\text{RH-D}}$\xspace}
\newcommand{\TRH}{$\text{T}_{\text{RH}}$\xspace}
\newcommand{\PMQTH}{$\text{T}_{\text{PMQ}}$\xspace}

\newcommand{\TRHDBASE}{$\widehat{\text{T}}_{\text{RH-D}}$\xspace}
\newcommand{\ABOACTQ}{$\text{ABO}_{\text{ACT}}(Q)$\xspace}

\newcommand{\ACTCTR}{\textit{ACTCTR}}

\newcommand{\blue}[1]{{\color{black}#1}}
\newcommand{\camready}[1]{{\color{purple}#1}}
\newcommand{\topic}[1]{\noindent\textbf{#1:}}
\newcommand{\DEFENSE}{$\text{PrISM}$\xspace}
\newcommand{\DEFENSEF}{$\text{PrISM-F}$\xspace}
\newcommand{\DEFENSED}{$\text{PrISM-D}$\xspace}
\newcommand{\DEFENSEPLUS}{$\text{PrISM+}$\xspace}
\newcommand{\ignore}[1]{}
\DeclareRobustCommand\encircle[1]{\tikz[baseline=(char.base)]{\node[shape=circle,fill,inner sep=1pt] (char) {\textcolor{white}{#1}}}}
%%%%%%% -- HELPER DEFINES ENDS -- %%%%%%%%%%%

%%%%%% -- PAPER CONTENT STARTS-- %%%%%%%%
\begin{abstract}
DRAM scaling has exacerbated the RowHammer vulnerability. To counter this, JEDEC recently introduced Per Row Activation Counting (PRAC) with the Alert Back-Off protocol as an optional DDR5 feature. While promising, PRAC requires per-row counter cells that incur area overhead, and updating them on every activation lengthens DRAM timing parameters, degrading performance. Probabilistic mitigations such as MINT offer a lower-cost alternative by randomly selecting and mitigating rows within periodic mitigation windows. MINT is effective at higher thresholds ($\geq\!1000$), but at lower thresholds, it must raise its mitigation rate to overcome the non-selection problem, where heavily hammered rows can repeatedly escape sampling. This fixed-rate scaling reduces effective memory bandwidth even when no attack is present.

To overcome this limitation, we propose PrISM, an intersection-based probabilistic mitigation that correlates sampled rows across windows using a Sampled History Queue (SHQ). PrISM samples a few activation slots per window, stores sampled-but-unmitigated rows in the SHQ, and requests an additional mitigation through the existing Alert Back-Off protocol when a sampled row reappears in this history. This allows PrISM to increase mitigation only when persistent row activity is observed, without globally increasing the fixed mitigation rate. At the threshold of 500, PrISM incurs a negligible 0.2\% average slowdown compared to 14\% for PRAC, with no DRAM array changes or per-row counters and only 625B of SRAM per bank, one to two orders of magnitude less than prior secure counter-based in-DRAM defenses. Compared to MINT, PrISM provides better scalability at low thresholds, reducing average slowdown from 10.7\% to 1.5\% at a threshold of 250, a 7.1$\times$ reduction. PrISM is open-sourced at \url{https://github.com/STAR-Laboratory/prism}.
\end{abstract}

\section{Introduction}
DRAM technology scaling has enabled large main memory capacities for data-intensive workloads. At the same time, smaller cells and tighter noise margins have made DRAM more vulnerable to disturbance effects. The most prominent example is the RowHammer vulnerability: a read-disturb phenomenon where repeatedly activating a DRAM row induces bit flips in physically adjacent rows~\cite{kim2014architectural, kim2014flipping}. Over the past decade, RowHammer has become a serious security problem, with several attacks compromising real systems~\cite{de2021smash, kwong2020rambleed, prisonbreak, lin2026gpubreach}. Meanwhile, aggressive scaling has sharply reduced the RowHammer threshold (\TRH{}), the minimum number of activations needed to induce bit flips. In this work, we focus on the double-sided RowHammer threshold (\TRHD{}), where two aggressor rows adjacent to a victim are alternately hammered.

\begin{figure}[b!]
\vspace{-0.1in}
\centering
\includegraphics[width=0.9\linewidth,height=\paperheight,keepaspectratio]{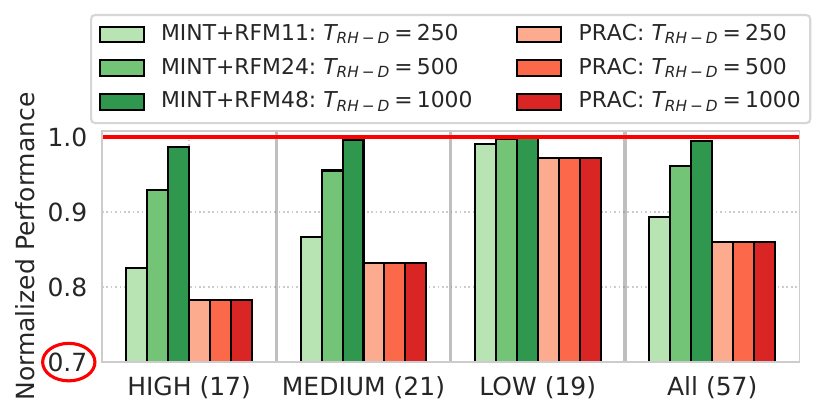}
\vspace{-0.1in}
\caption{Performance overhead of PRAC~\cite{qprac} and MINT~\cite{MINT} under varying double-sided RowHammer thresholds (\TRHD{}). On high-memory-intensity workloads, PRAC incurs an average 21.8\% slowdown due to updates to the activation counter on every activation. In contrast, MINT incurs only 1.4\% overhead at \TRHD{} of 1000, but its slowdown increases to 17.5\% at \TRHD{} of 250 as lower thresholds require more frequent mitigations.}
% \vspace{-0.05in}
\label{fig:motivation}
\end{figure}

To counter RowHammer, JEDEC standardized Per Row Activation Counting (PRAC) for DDR5~\cite{jedec_ddr5_prac}. PRAC maintains a per-row activation counter and uses the \emph{Alert Back-Off (ABO)} protocol. When a counter reaches the Back-Off threshold, DRAM asserts \ALERT{} and the controller issues a Refresh Management (RFM) command while pausing regular requests for a fixed interval (e.g., 350ns). Recent designs such as QPRAC~\cite{qprac} and MOAT~\cite{qureshi2024moat} show that PRAC can provide strong protection even at sub-100 \TRH{}, but at high cost~\cite{Chronus,mopac_isca25}. Because PRAC updates a per-row counter on every activation through a read-modify-write operation, it increases the \TRP{} and \TRC{} timing parameters, resulting in significant performance degradation. As \cref{fig:motivation} shows, PRAC, implemented using QPRAC, incurs 14\% average slowdown across all workloads and 21.8\% slowdown on high-memory-intensity workloads ($\geq$10 row-buffer misses per kilo-instruction), largely independent of \TRHD{} (see \cref{sec:eval_methodology} for methodology). PRAC also incurs notable area overhead due to its per-row counters~\cite{DSAC, MINT}. Since PRAC is an \emph{optional} DDR5 feature~\cite{jedec_ddr5_prac} and its commercial adoption is uncertain given these costs, there is a clear need for practical low-overhead mitigations for near-term DDR5 systems~\cite{techpowerup_ddr6_2027}.

Probabilistic in-DRAM mitigations offer a much lower-cost alternative to PRAC~\cite{PROTEAS, MINT, jaleel2024pride,APT}. These schemes avoid per-row counters by randomly mitigating a subset of recently activated rows using either Target Row Refresh (TRR), which borrows time from periodic refreshes~\cite{hassan2021UTRR, JEDEC-DDR4}, or \RFM{}, which temporarily blocks normal memory service while DRAM performs the required mitigation. Unless otherwise stated, we assume one TRR mitigation opportunity every two \TREFI{} intervals. Because each \RFM{} consumes memory bandwidth and delays regular requests, probabilistic schemes are most effective when RFMs are issued only occasionally. This works well at higher thresholds (\TRHD{} $\!\geq\!$ 1000), where infrequent mitigations are enough to provide security. For example, MINT~\cite{MINT} requires mitigations only every 48 activations at \TRHD{} of 1000, resulting in only a 1.4\% slowdown on high-memory-intensity workloads. This makes MINT highly attractive at sufficiently high thresholds, especially given its extremely small in-DRAM storage cost.

At lower thresholds, however, probabilistic schemes face a statistical barrier: the \textbf{non-selection problem}. Each mitigation window mitigates at most one randomly chosen row, so a heavily hammered aggressor may be skipped across many windows and remain unmitigated long enough to induce bit flips. To maintain security, MINT \emph{statically} increases its mitigation rate, issuing RFMs more frequently even when no aggressor is present. This raises overhead as \TRHD{} drops, especially for high-memory-intensity workloads. As \cref{fig:motivation} shows, MINT performs mitigations every 24 activations at \TRHD{} of 500, increasing slowdown to 7.1\%, and every 11 activations at \TRHD{} of 250, increasing slowdown to 17.5\%. These results motivate a probabilistic defense that scales to lower thresholds without uniformly increasing mitigation frequency.

To this end, we propose \textbf{\DEFENSE{}}, \underline{Pr}obabilistic \underline{I}ntersection-based \underline{S}ampling \underline{M}itigation. The key insight is that rows that are repeatedly activated are more likely to reappear across mitigation windows, whereas benign rows rarely reappear in the sampled history. \DEFENSE{} exploits this temporal correlation to address the \emph{non-selection problem}. Instead of statically raising the mitigation rate, \DEFENSE{} samples a small set of activation slots in each window and tracks their row addresses across a recent \emph{lookback window}. When a newly sampled row matches a row already in this history, \DEFENSE{} requests an \emph{additional} mitigation through the existing ABO protocol.

This mechanism allows \DEFENSE{} to increase mitigation only when persistent row activity is observed, without scaling the default mitigation rate. In MINT, a mitigation window with $W$ activation slots selects only one row, so a persistent aggressor is selected with probability $1/W$, about 1.4\% when $W\!=\!72$. \DEFENSE{} instead samples $R$ slots per window and remembers sampled-but-unmitigated rows for $L$ windows using the Sampled History Queue (SHQ). Thus, a row that appears in every window has probability $R/W$ of being sampled in each window, and its chance of appearing at least once in the lookback history is roughly $1-(1-R/W)^L$. With $W\!=\!72$, $R\!=\!7$, and $L\!=\!41$, this probability exceeds 98\%, making a persistent aggressor highly likely to create an SHQ intersection while keeping the default mitigation rate low.

Crucially, \DEFENSE{} requests few additional \ALERT{}-induced \RFM{}s for benign applications. A benign row accessed only occasionally is unlikely to be sampled repeatedly within the SHQ lookback window, so it seldom creates intersections and therefore rarely triggers additional mitigations. Thus, \DEFENSE{} preserves the counter-free nature of probabilistic mitigations while requesting additional mitigations only for repeated sampled activity. In contrast, MINT maintains security at low \TRHD{} by increasing the mitigation rate for all workloads, reducing effective memory bandwidth.

\DEFENSE{} is fully compatible with the existing JEDEC ABO protocol and requires no DRAM array or interface changes. Compared to PRAC, \DEFENSE{} avoids per-row counters and counter updates on every activation, which lengthen DRAM timing parameters and cause 14\% average slowdown. Compared to MINT, \DEFENSE{} improves scalability at low \TRHD{} by keeping the default mitigation rate low and requesting additional mitigations only when SHQ intersections indicate repeated sampled activity. At \TRHD{} of 500, \DEFENSE{} achieves a negligible 0.2\% average slowdown while requiring only 625B of SRAM per bank, which is about 20$\times$ and 170$\times$ smaller than prior secure counter-based in-DRAM mitigations such as Mithril~\cite{kim2022mithril} and ProTRR~\cite{ProTRR}, respectively. Even at an ultra-low \TRHD{} of 250, where MINT incurs 10.7\% average slowdown due to frequent fixed-rate mitigations, \DEFENSE{} incurs only 1.5\% average slowdown, a 7.1$\times$ reduction.

\smallskip
\noindent\textbf{Summary of Contributions:} 
\begin{itemize}[leftmargin=*, itemsep=3pt, topsep=2pt]
    \item We identify the \textbf{non-selection problem} as the key barrier for fixed-rate probabilistic RowHammer defenses at low \TRHD{}, and propose \textbf{\DEFENSE{}}, which addresses it by correlating sampled row addresses across windows.

    \item \DEFENSE{} uses SHQ intersections to request additional mitigations only for repeated sampled activity, keeping the default \RFM{} rate low. It reuses the existing ABO protocol and requires no changes to the DRAM array or interface.

    \item \DEFENSE{} avoids PRAC's costly per-row counters and counter updates on each activation. At \TRHD{} of 500, \DEFENSE{} achieves a negligible 0.2\% average slowdown, compared to 14\% for PRAC, while requiring only 625B of per-bank SRAM.

    \item We show that \DEFENSE{} improves low-threshold scalability over MINT by avoiding the need to uniformly increase RFM frequency. On high-memory-intensity workloads, \DEFENSE{} reduces slowdown from 7.1\% to 0.5\% at \TRHD{} of 500 and from 17.5\% to 2.5\% at \TRHD{} of 250.
\end{itemize}
\section{Background and Motivation}\label{sec:background_motivation}
\subsection{Threat Model}\label{subsec:threat_model}
We consider a DRAM-based system vulnerable to RowHammer. The adversary is unprivileged, knows the deployed in-DRAM defenses, and can craft tailored access patterns to bypass them~\cite{meyer2026phoenix,jattke2021blacksmith,marionette_asplos25}. In our evaluation, all probabilistic mitigations, including MINT and \DEFENSE{}, use fractal mitigation~\cite{autorfm_hpca25} to defend against transitive attacks~\cite{HalfDouble}.

We focus on activation-driven RowHammer. RowPress~\cite{rowpress} and ColumnDisturb~\cite{columndisturb_micro25} are outside our primary threat model; Appendix~\ref{app:rowpress_columndisturb} discusses possible extensions.

\subsection{The RowHammer Vulnerability}
RowHammer is a read-disturbance phenomenon in which repeated activations of aggressor rows disturb charge in nearby victim rows and can induce bit flips~\cite{kim2014flipping, tuugrul2025understanding, kim2025sok, DSN23_PTGuard, citadel_rh_micro25}. The RowHammer threshold (\TRH{}), the minimum number of activations needed to induce a bit flip, has decreased significantly across DRAM generations~\cite{kim2020revisitingRH}. We use the double-sided threshold (\TRHD{}), where two aggressors adjacent to a victim are alternately hammered, as our primary metric.

RowHammer is both a security and a reliability concern. It has enabled attacks across CPUs and GPUs~\cite{vanderveen2016drammer, lin2025gpuhammer, hu2026gddrhammer}, including privilege escalation~\cite{seaborn2015exploiting, lin2026gpubreach}, ML model degradation~\cite{hong2019terminal, deephammer}, and other exploits~\cite{sgx-bomb, gruss2016rhjs, cojocar2019eccploit, gruss2018another, flipfengshui, prisonbreak, SALT}. Recent work also shows that RowHammer mitigations can introduce denial-of-service and timing-channel attack surfaces~\cite{dapper, prac_timing_channel_isca25, prac_tc_micro}. Bit flips have even been observed under benign workloads~\cite{loughlin2022moesi}, further underscoring the need for secure and low-overhead RowHammer defenses.

\subsection{Target Row Refresh and Refresh Management}
Commodity DRAM has relied primarily on two in-DRAM RowHammer mitigation methods:

\smallskip
\noindent\textbf{Target Row Refresh (TRR)}: 
TRR tracks potential aggressor rows using small per-bank trackers, typically with 4--28 entries, or probabilistic sampling~\cite{jattke2021blacksmith, ecc_fail, hassan2021UTRR}. It then refreshes nearby victim rows within a fixed blast radius (e.g., $\text{BR}\!=\!2$) during periodic refreshes by borrowing time from the refresh cycle (\TRFC{})\footnote{We assume one TRR opportunity per two \TREFI{} intervals, based on prior studies~\cite{ecc_fail, hassan2021UTRR}. \cref{subsec:results-senstrr} evaluates sensitivity to this rate.}~\cite{olgun2024HBM2study, hassan2021UTRR}. However, TRR's limited tracking capacity makes it vulnerable to crafted access patterns, and prior works have bypassed TRR on DDR4~\cite{jattke2021blacksmith, HalfDouble, frigo2020trrespass} and DDR5 devices~\cite{jattke2024zenhammer, meyer2026phoenix}.

\smallskip
\noindent\textbf{Refresh Management (\RFM{})}: 
To address TRR's limitations, DDR5 introduced RFM~\cite{jedec_ddr5_prac}. The memory controller monitors per-bank activations and issues an RFM once a threshold is reached, allowing DRAM to perform RowHammer mitigations. DDR5 defines two RFM types: All-Bank RFM (\RFMAB{}), which applies mitigations across all banks, and Same-Bank RFM (\RFMSB{}), which applies mitigations to the same bank ID across all bank groups (e.g., Bank~0 in every bank group).

\subsection{Per Row Activation Counting (PRAC)}\label{subsec:PRAC}
To address the worsening RowHammer vulnerability, JEDEC standardized \textit{Per Row Activation Counting (PRAC)} in the DDR5 specification~\cite{jedec_ddr5_prac}. PRAC enables precise aggressor tracking through two primary mechanisms.

\smallskip
\noindent\textbf{Per-Row Activation Counters}: 
Each DRAM row is equipped with counter cells~\cite{bennett2021panopticon}. On every activation, the corresponding counter is incremented via a read--modify--write during precharge, increasing core timings such as \TRP{} and \TRC{}.

\smallskip
\noindent\textbf{Alert Back-Off (ABO) Protocol}: 
PRAC also defines the Alert Back-Off (ABO) Protocol, which lets DRAM request mitigation time from the memory controller. When a counter or mitigation queue reaches the Back-Off threshold (\NBO{}), DRAM asserts \ALERT{}; after up to 180ns during which the controller may continue issuing regular DRAM commands, the controller issues a configured number of \RFMAB{} commands (\NMIT{} $\in \{1,2,4\}$). Each \RFMAB{} stalls the entire channel for 350 ns, giving DRAM time to perform targeted refreshes. The current specification also requires an \ABODELAY{} of \NMIT{} activations before a subsequent \ALERT{}.

\smallskip
Recent studies, including QPRAC~\cite{qprac} and MOAT~\cite{qureshi2024moat}, propose secure PRAC implementations that demonstrate robust RowHammer protection even at sub-100 \TRH{}~\cite{Chronus}. In this paper, we use QPRAC as our baseline PRAC design.

\subsection{Pitfall of PRAC: Performance and Area Overhead}
Secure PRAC designs can provide strong protection even at sub-100 \TRH{}, but they suffer from two key drawbacks.

\smallskip
\noindent\textbf{Performance Overhead}:
PRAC updates a per-row activation counter on every activation through a read--modify--write operation during precharge~\cite{jedec_ddr5_prac}. This increases key timing parameters: \TRP{} rises from 16ns to 36ns (2.25$\times$), and \TRC{} rises from 48ns to 52ns~\cite{mopac_isca25}. As a result, row-buffer-conflict latency increases by 42\%, while the longer \TRC{} reduces the per-bank activation rate by 8\%. Prior work reports an average slowdown of $\sim$10\% due to PRAC timing overheads~\cite{mopac_isca25, Chronus}.

\begin{figure}[h!]
\centering
\vspace{-0.05in}
\includegraphics[width=0.9\linewidth,height=\paperheight,keepaspectratio]{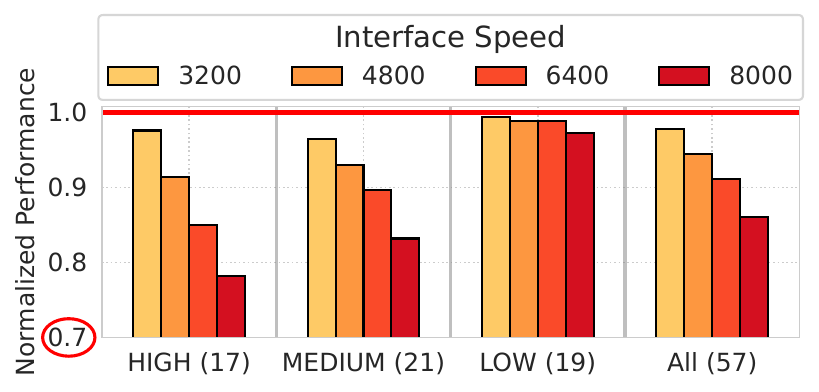}
\vspace{-0.1in}
\caption{Normalized performance of PRAC~\cite{qprac} across DDR5 interface speeds. At 3200 MT/s, PRAC incurs a 2.2\% slowdown, which rises to 14\% at 8000 MT/s due to amplified timing penalties at higher data rates.}
\vspace{-0.05in}
\label{fig:motivation_interface}
\end{figure}

We further show that PRAC's performance overhead grows sharply with interface speed. \Cref{fig:motivation_interface} shows PRAC performance across DDR5 data rates. PRAC incurs a 2.2\% average slowdown at 3200 MT/s, but this rises to 14\% at 8000 MT/s, where shorter tRRD and tFAW allow the controller to issue activations across banks more rapidly, exposing PRAC's fixed timing penalties more often\footnote{Appendix~\ref{app:prac-real-system} provides a detailed comparison with recent real-system PRAC performance measurements~\cite{prac_perf_cal}.}. As DRAM data rates continue to scale (e.g., up to 17.6Gb/s in DDR6~\cite{techpowerup_ddr6_2027}), these timing penalties may become more significant.

\noindent\textbf{Area Overhead}: 
PRAC also incurs notable area overhead because each row requires additional counter cells and update logic. An early Samsung DSAC work~\cite{DSAC} estimates that these additions increase overall DRAM core area by roughly 9\%, which is significant for the price-sensitive DRAM market.

PRAC is currently defined as an \emph{optional} feature in the DDR5 specification~\cite{jedec_ddr5_prac}. DDR5 is expected to remain the dominant DRAM technology for several years before DDR6, where PRAC is expected to see broader adoption and may become mandatory, as discussed at DRAMSec'24~\cite{techpowerup_ddr6_2027}. Developing practical, low-overhead RowHammer mitigations for current and near-term DDR5 systems is therefore critical.

\subsection{Scalability Limits of Probabilistic Mitigations}
Prior work has proposed lightweight probabilistic defenses such as PrIDE~\cite{jaleel2024pride} and MINT~\cite{MINT}, which require only a small in-DRAM state. MINT randomly selects one activation slot per \emph{mitigation window}, defined as the maximum number of activations between mitigations, and mitigates the corresponding row at the window's end. At \TRHD{} $\!\geq\!$ 1000, this design provides strong protection with negligible slowdown.

At lower \TRHD{}, however, probabilistic defenses face a fundamental \emph{non-selection problem}: random sampling can miss an aggressively hammered row across multiple windows, allowing it to remain unmitigated long enough to induce bit flips. MINT maintains security at lower thresholds by statically increasing its mitigation rate and issuing RFMs more frequently, even when aggressive behavior is absent. Each \RFM{} temporarily blocks memory requests during mitigation, reducing effective DRAM bandwidth. For example, MINT issues an RFM every 24 activations at \TRHD{} of 500 and every 11 activations at \TRHD{} of 250, reducing available bandwidth by nearly 23\% and 40\%. On high-memory-intensity workloads, this causes 7.1\% and 17.5\% slowdown, respectively. Thus, while MINT remains highly storage-efficient, its fixed-rate scaling becomes increasingly costly at low \TRHD{}. Our goal is to design a scalable probabilistic mitigation that preserves high performance even at low thresholds.
\section{PrISM: A Scalable Probabilistic Mitigation}\label{sec:prism}
We propose \DEFENSE{}, \underline{Pr}obabilistic \underline{I}ntersection-based \underline{S}ampling \underline{M}itigation, a scalable probabilistic RowHammer mitigation that solves the \emph{non-selection problem}. Fixed-rate probabilistic defenses such as MINT~\cite{MINT} provide strong protection and require extremely small in-DRAM state, but at low thresholds (\TRHD{}\! $\leq$\! 500), they must issue RFMs frequently even when the system is not under attack, reducing effective memory bandwidth and degrading performance.

To solve this, \DEFENSE{} correlates sampled row addresses across mitigation windows and requests \emph{additional} mitigations through the Alert Back-Off (ABO) protocol only when repeated sampled activity is observed. In each mitigation window, \DEFENSE{} samples multiple activation slots and compares the sampled rows against a bounded history of previously sampled rows. We define an \emph{intersection} as a match between a newly sampled row and a row in this history. A heavily activated row is more likely to reappear across windows and create intersections, making it eligible for additional mitigations. In contrast, benign rows accessed irregularly are less likely to create intersections, avoiding unnecessary mitigations. 

This allows \DEFENSE{} to use a larger default window $W$ than MINT, reducing the default RFM rate while preserving strong RowHammer protection. As a result, \DEFENSE{} scales to lower \TRHD{} without uniformly increasing mitigation frequency, preserving performance relative to fixed-rate probabilistic defenses. Compared to PRAC, \DEFENSE{} avoids DRAM array changes, per-row counters, and counter updates on every activation, reducing both performance and area overhead.

\begin{figure}[b!]
    % \vspace{-0.1in}
    \centering    
    \includegraphics[width=0.95\linewidth,height=\paperheight,keepaspectratio]{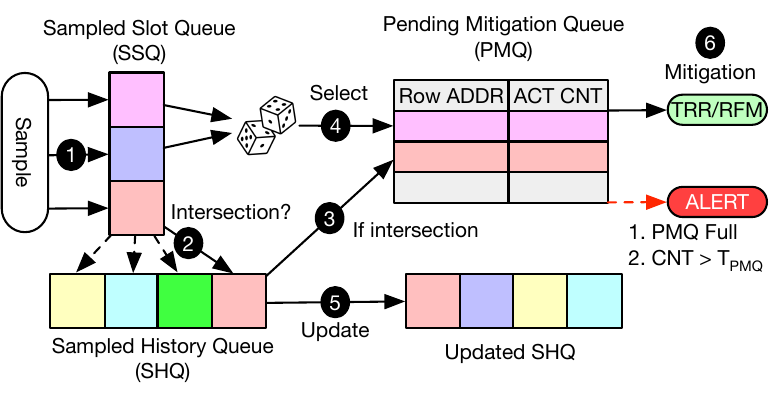}
    % \vspace{-0.1in}   
    \caption{Design and operation of \DEFENSE{}. In each mitigation window, \DEFENSE{} samples a few activation slots into the Sampled Slot Queue (SSQ) and checks them against the Sampled History Queue (SHQ), which stores sampled-but-unselected rows from recent windows. Rows that intersect with the SHQ are enqueued into the Pending Mitigation Queue (PMQ) for additional mitigation. \DEFENSE{} also randomly selects one non-intersecting sampled row for the PMQ as the default mitigation candidate, while inserting the remaining sampled-but-unselected rows into the SHQ. For each default mitigation opportunity (TRR or RFM), \DEFENSE{} mitigates and removes the highest-count PMQ entry. If the PMQ becomes full or any entry's count exceeds \PMQTH{}, \DEFENSE{} asserts \ALERT{} to request an additional mitigation.}
    \label{fig:prism-overview}
    % \vspace{-0.1in}
\end{figure}

\subsection{Design of PrISM}
\DEFENSE{} enables intersection-based additional mitigations using three per-bank structures, as shown in~\cref{fig:prism-overview}. The \textbf{Sampled Slot Queue (SSQ)} buffers row addresses of sampled and activated slots, from which \DEFENSE{} selects the default mitigation candidate at the end of each window. 
The \textbf{Sampled History Queue (SHQ)} retains sampled-but-unselected rows from the previous $L$ windows, called the lookback window, and serves as the history for intersection checks. When a newly sampled row matches an SHQ entry, \DEFENSE{} enqueues the intersecting row into the \textbf{Pending Mitigation Queue (PMQ)} for additional mitigation. \DEFENSE{} also enqueues one randomly selected non-intersecting sample as the default mitigation candidate, so each window still contributes one default probabilistic mitigation. The remaining sampled-but-unselected rows are inserted into the SHQ for future intersection checks.

The PMQ buffers rows selected for mitigation but not yet serviced. Each entry contains a row address and an activation counter that tracks activations while the row resides in the PMQ. At each mitigation opportunity, \DEFENSE{} mitigates the highest-count entry and dequeues it; if the PMQ becomes full or any entry's counter exceeds the \emph{tardiness threshold} (\PMQTH{}), \DEFENSE{} requests additional mitigation through the existing ABO protocol. Unless otherwise stated, we use a 16-entry PMQ and set \PMQTH{} to 4. 

By decoupling mitigation selection from mitigation service, the PMQ provides three benefits. First, it makes \DEFENSE{} compatible with refresh and \RFM{} postponement~\cite{jedec_ddr5_prac}, since selected rows can wait until a mitigation opportunity becomes available. Second, because \ALERT{} is observed at the channel level while banks complete windows at different times, per-bank PMQs let other banks retain useful pending entries when any bank raises \ALERT{}; an ABO-triggered RFM can then service pending rows across banks, reducing future \ALERT{}s. Third, intersecting rows may be serviced by default TRR/RFM opportunities before \ALERT{} becomes necessary, further reducing the \ALERT{} frequency and improving performance.

\subsection{Operation of PrISM}
Like MINT, \DEFENSE{} operates on fixed-size mitigation windows. However, \DEFENSE{} can use a longer default window $W$ because it requests additional mitigations when repeated sampled activity creates intersections. \cref{fig:prism-overview} illustrates the operation of \DEFENSE{}.
\encircle{1} At the beginning of each mitigation window, \DEFENSE{} randomly samples $R$ activation slots out of the $W$ slots and records each sampled-and-activated row in the SSQ.
\encircle{2} When a sampled activation occurs, \DEFENSE{} checks whether the activated row matches any SHQ entry.
\encircle{3} If the row produces an intersection, \DEFENSE{} enqueues it into the PMQ for additional mitigation. 
\encircle{4} At the end of the window, \DEFENSE{} randomly selects one non-intersecting sampled row from the SSQ and enqueues it into the PMQ as the default mitigation candidate. This ensures that each window contributes one default probabilistic mitigation, regardless of whether intersections occur.
\encircle{5} \DEFENSE{} then updates the SHQ in FIFO order. Up to $R\!-\!1$ sampled-but-unselected rows from the current window are enqueued, while the oldest entries are evicted. If fewer than $R\!-\!1$ such rows exist because some sampled slots were never activated, \DEFENSE{} inserts invalid placeholders to keep the lookback window $L$ deterministic.
\encircle{6} At each default mitigation opportunity, such as a TRR during a refresh or an RFM, \DEFENSE{} mitigates and removes the PMQ entry with the highest activation count. If the PMQ becomes full or any entry's counter exceeds \PMQTH{}, \DEFENSE{} requests an additional mitigation via the Alert Back-Off (ABO) protocol, again servicing the entry with the highest counter.

Under benign workloads, intersections are infrequent because most rows are accessed sparsely and irregularly~\cite{saxena2024rubix, saileshwar2022RRS, qprac, dapper}. In these common cases, the PMQ is mainly populated by default mitigation candidates, which are drained by upcoming default mitigations. \DEFENSE{} therefore issues few \ALERT{}s and incurs negligible performance overhead.

\subsection{JEDEC-Compatible Operation of PrISM}\label{subsec:prism-jedec}
\DEFENSE{} uses the existing JEDEC ABO protocol to request additional mitigations with \emph{one} RFM per \ALERT{}. The key challenge is that \ALERT{} does not immediately drain a pending mitigation: after \ALERT{}, up to three activations (\ABOACT{}) can occur before the corresponding \RFM{}, and one \ABODELAY{} activation is required before \ALERT{} can be reasserted. Thus, under one RFM per \ALERT{}, ABO can drain only one pending entry every four activations, while the worst case can produce one intersection per activation. \DEFENSE{} handles this by sizing the Sampled Slot Queue (SSQ) to temporarily buffer intersecting samples until PMQ entries are freed.

\begin{figure}[h!]
    % \vspace{-0.1in}
    \centering    
    \includegraphics[width=0.95\linewidth,height=\paperheight,keepaspectratio]{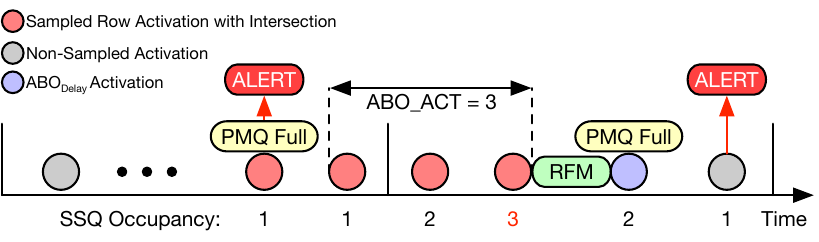}
    % \vspace{-0.1in}   
    \caption{Sampled slots can cluster near adjacent-window boundaries, and all intersect with the SHQ, producing intersections faster than ABO can drain. The SSQ buffers intersecting samples that cannot yet enter the PMQ.}
    \label{fig:ssq-size}
    % \vspace{-0.1in}
\end{figure}

\cref{fig:ssq-size} illustrates this worst case for $R\!=\!2$. Since \DEFENSE{} samples $R$ slots uniformly at random per window, all $R$ slots can cluster near the boundary of two adjacent windows, and all may intersect with the SHQ, producing up to $2R$ intersections in $2R$ consecutive activations. The first intersection enters the PMQ and triggers \ALERT{}; the remaining intersections wait in the SSQ until ABO frees PMQ slots. Because ABO drains one pending entry every four activations, the SSQ must cover the peak number of waiting intersections during the burst. Of the $2R\!-\!1$ intersections after the first, $\lfloor(2R-1)/4\rfloor$ can be drained before the burst completes. Thus, the required SSQ size is:
\begin{equation}
    S_{\mathrm{SSQ}} \geq
    (2R - 1) -
    \left\lfloor \frac{2R - 1}{4} \right\rfloor
\end{equation}
For our largest evaluated $R\!=\!9$, this bound requires 13 SSQ entries. Beyond a single boundary burst, repeated bursts remain bounded as long as the long-run intersection rate does not exceed the ABO drain rate. Since at most $R$ samples can intersect per $W$-activation window, $W\geq4R$ is sufficient.

\subsection{Address Mapping for Ultra-Low \TRHD{}}\label{subsec:prism-rand-mapping}
At ultra-low \TRHD{} ($\leq\!250$), even benign workloads can create frequent intersections because conventional mappings preserve spatial locality. For example, Minimalist Open-Page (MOP) mapping~\cite{MOP} places nearby cache lines in the same DRAM row, improving row-buffer locality but also causing the same row to be sampled across multiple windows, reappear in the SHQ, and trigger unnecessary \ALERT{}s.

To reduce this overhead, \DEFENSE{} uses randomized address mapping only for ultra-low \TRHD{} configurations. Randomized mapping spreads nearby cache lines across many DRAM rows, reducing persistent row-level locality. We use Rubix-style randomization~\cite{saxena2024rubix}, following prior work~\cite{autorfm_hpca25}. \Cref{tab:random_mapping_alerts} shows the average per-channel \ALERT{} rate under both mappings. Randomization reduces benign \ALERT{}s across all thresholds, with the largest absolute reduction at \TRHD{} $=250$, where it cuts the \ALERT{} rate from 312.3 to 50.3 per 1K \TREFI{} ($6.2\times$). Thus, \DEFENSE{} enables randomized mapping only at ultra-low \TRHD{} ($\leq250$), while retaining conventional mapping at higher thresholds, where \ALERT{} rates are already low and preserving row-buffer locality is preferable.

\begin{table}[h!]
\centering
\caption{Average per-channel \ALERT{} rate of \DEFENSE{}}
\begin{footnotesize}
\resizebox{0.9\columnwidth}{!}{%
\begin{tabular}{cccc}
\toprule
\textbf{Target \TRHD{}} &
\textbf{\shortstack{MOP\\\ALERT{}s / 1K \TREFI{}}} &
\textbf{\shortstack{Random\\\ALERT{}s / 1K \TREFI{}}} &
\textbf{Reduction} \\
\midrule
1000 & 2.4   & 0.04  & 60$\times$ \\
500  & 32.2  & 3.5  & 9.2$\times$ \\
\textbf{250}  & \textbf{312.3} & \textbf{50.3} & \textbf{6.2$\times$} \\
\bottomrule
\end{tabular}
}
\end{footnotesize}
\label{tab:random_mapping_alerts}
\end{table}

\section{Security Analysis of PrISM}\label{sec:security_analysis}
We analyze the security of \DEFENSE{} against an attack pattern that maximizes the number of unmitigated activations to any aggressor row within a 32ms refresh window (\TREFW{}). We quantify security using the \emph{Mean-Time-to-Failure (MTTF)} metric. Unless otherwise stated, we target a per-bank MTTF of 10,000 years, comparable to the intrinsic DRAM soft-error rate~\cite{ddr4_error_rate}, consistent with prior work~\cite{MINT, jaleel2024pride, autorfm_hpca25, mopac_isca25}. 

\subsection{Worst-Case Attack: Circular-$X$-Rows Attack}\label{subsec:worst_case}
The attacker must balance two competing goals. First, it wants to maximize the number of activations delivered to aggressor rows within \TREFW{}. Second, it wants each aggressor to avoid \DEFENSE{}'s mitigations, including the default mitigation applied in every window and the additional intersection-based mitigations provided by SHQ. Repeated sampling increases the chance that a row is mitigated by the default mitigation or creates intersections that make the row eligible for additional mitigation. The worst-case pattern for \DEFENSE{} is therefore the pattern that uses the full activation budget within \TREFW{} while spreading activations across enough rows to minimize avoidable default selections and intersections.

We model this worst case using a \emph{Circular-$X$-rows attack}. The attacker chooses $X$ aggressor rows and repeatedly activates them in a round-robin manner, one row per activation slot. For a mitigation window of $W$ activation slots, the row activated at slot $s$ of window $t$ is:
% \vspace{-0.05in}
\begin{equation}
    \text{row}(t,s) = (tW + s) \bmod X
\label{eq:circular_attack}
\end{equation}
The single parameter $X$ controls the entire attack spectrum, exposing a fundamental trade-off between hammering rate per aggressor and evasion of SHQ intersections. Sweeping $X$ over $[W, (L+1)W]$ captures all meaningful attacker strategies:

\begin{enumerate}[leftmargin=*, itemsep=4pt, topsep=5pt]
    \item \textbf{$X = W$:} Every window contains $W$ distinct rows hammered once each. This maximizes the activation rate per aggressor, but it also maximizes exposure to \DEFENSE{} because the same rows repeatedly appear within the SHQ lookback window. This corresponds to the multi-row, single-copy pattern analyzed by MINT~\cite{MINT}.
    
    \item \textbf{$W < X < (L+1)\,W$:} The attacker spreads activations across more aggressor rows. This reduces the chance that a row reappears within the $L$-window SHQ history and therefore lowers the probability of intersections. However, it also reduces the number of activations delivered to each aggressor within \TREFW{}. Thus, intermediate $X$ values capture the fundamental tradeoff between hammering rate and mitigation exposure.
    
    \item \textbf{$X = (L+1)\,W$:} The target row reappears exactly at the edge of the lookback window, completely evading SHQ intersections. Any larger $X$ also evades intersections but reduces per-aggressor hammering rate by a factor of $\frac{X}{(L+1)W}$, weakening the attack. We therefore do not sweep beyond $X=(L+1)W$.
\end{enumerate}

Three properties establish why the circular-$X$-rows pattern captures the worst case for \DEFENSE{}.

\smallskip\topic{1. Intra-window ordering is irrelevant} Because \DEFENSE{} samples $R$ slots uniformly at random from the $W$ slots of a window, a row's sampling probability depends only on how many times it appears in the window, not on where those appearances fall~\cite{MINT}. Reordering activations within a window cannot change the probability of default selection, SHQ insertion, or intersection.

\smallskip\topic{2. Using one copy per row per window is worst-case} If a row appears $c$ times within a single window, its probability of being sampled at least once is:
% \vspace{-0.05in}
\begin{equation}
    P_{\text{sample}}(c) = 1 - \frac{\binom{W-c}{R}}{\binom{W}{R}}
\label{eq:sample_c}
\end{equation}
$P_\text{sample}(c)$ increases with $c$, making the row more likely to be selected by the default mitigation, inserted into the SHQ, or causing additional mitigations through intersections. Additional copies of the same row also consume activation slots that could instead be used to attack other rows. Thus, to maximize the number of independent aggressors while minimizing avoidable sampling exposure, the attacker should activate each aggressor at most once per window.

\smallskip\topic{3. Round-robin spacing is the strongest single-copy schedule} Under the single-copy constraint, the attacker's only remaining choice is how frequently to activate each row across windows. Clustering repeated activations of the same row increases the chance that the row remains within the $L$-window SHQ history, creating more intersections. Spreading activations avoids intersections, but reduces the per-aggressor hammering rate. Neither direction beats the optimal stationary spacing, so a periodic round-robin is optimal in the worst case.

\subsection{Determining Minimum Supported T$_{\text{RH-D}}$ of PrISM}\label{subsec:trhd_analysis}
The security of \DEFENSE{} is determined by \emph{three} design parameters: the window size ($W$), the number of per-window sampled slots ($R$), and the lookback window ($L$). For each $(W, R, L)$, we sweep $X\!\in\![W,\, (L+1)W]$ and take the maximum required \TRHD{} as the security bound. 

\smallskip
\topic{Per-window mitigation probability}
For a circular-$X$-rows attack, let $K$ denote the number of prior appearances of the same aggressor that fall within the SHQ's $L$ lookback window (approximately $L\! \cdot\! W/X$). We model this SHQ residency probability with the following steady-state fixed point\footnote{{Appendix~\ref{app:detailed_security} provides the complete derivation for this occupancy model.}}: 
\begin{equation}
P_{\text{SHQ}} =
\frac{
K\left(R-1+\left(P_{\text{SHQ}}\right)^R\right)
}{
W + K \cdot R
}
\label{eq:pshq_k}
\end{equation}
Furthermore, \DEFENSE{} prioritizes rows \emph{not} already present in the SHQ for default mitigation. Thus, the default mitigation contributes a new mitigation only when at least one of the $R$ sampled rows is absent from the SHQ. The probability that all $R$ sampled rows are already present in the SHQ is approximately $(P_{\text{SHQ}})^R$. Combining the effective default mitigation with the intersection-based mitigation, the per-window mitigation probability is:
\begin{equation}
P_m =
\underbrace{
\frac{1-\left(P_{\text{SHQ}}\right)^R}{W}
}_{\text{default mitigation}}
+
\underbrace{
\frac{R}{W} \cdot P_{\text{SHQ}}
}_{\text{intersection-based eligibility}} 
\label{eq:pm_k}
\end{equation}

We then apply the Saroiu-Wolman model~\cite{Saroiu} to convert $P_m$ into the minimum \TRHD{} that meets our target MTTF (10,000 years), and sweep $X \in [W, (L\!+\!1)W]$ to find the worst-case $X^\ast$ for each $(W, R, L)$.

\begin{figure}[h!]
    % \vspace{-0.15in}
    \centering    
    \includegraphics[width=0.95\linewidth,height=\paperheight,keepaspectratio]{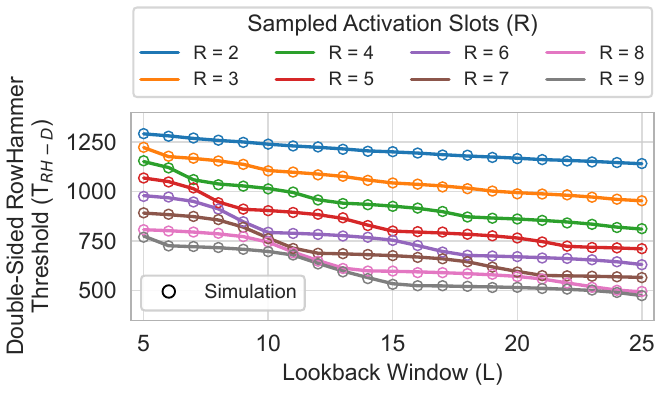}
    % \vspace{-0.15in}   
    \caption{Minimum supported \TRHD{} of \DEFENSE{} under the worst-case circular-$X$-rows attack at $W\!=\!72$, across sampled activation slots ($R$) and lookback windows ($L$). Larger $R$ and $L$ jointly increase the per-window mitigation probability, sharply lowering the supported \TRHD{}.}
    \label{fig:w72-trh}
    % \vspace{-0.1in}
\end{figure}

\smallskip\topic{Impact of $R$ and $L$ at fixed $W$}
\cref{fig:w72-trh} shows the supported \TRHD{} as we vary the number of sampled activation slots ($R$) and the SHQ lookback window ($L$), at a window size of $W\!=\!72$. Increasing either $R$ or $L$ improves security. A larger $R$ increases the probability of sampling an aggressor in each window, increasing the chance that any hammered row is mitigated by the default mitigation or SHQ intersection. Larger $L$ expands the history window, increasing the chance that an aggressor reappearing across windows produces an intersection. For example, \DEFENSE{} can support \TRHD{} of 954 with $R\!=\!3$ and $L\!=\!25$, while raising $R$ to 8 at the same $L$ brings the supported \TRHD{} down to 494. We validate our analytical model using a 5-million-epoch Monte Carlo simulation that closely matches it.

\begin{figure}[h!]
    % \vspace{-0.13in}
    \centering    
    \includegraphics[width=0.95\linewidth,height=\paperheight,keepaspectratio]{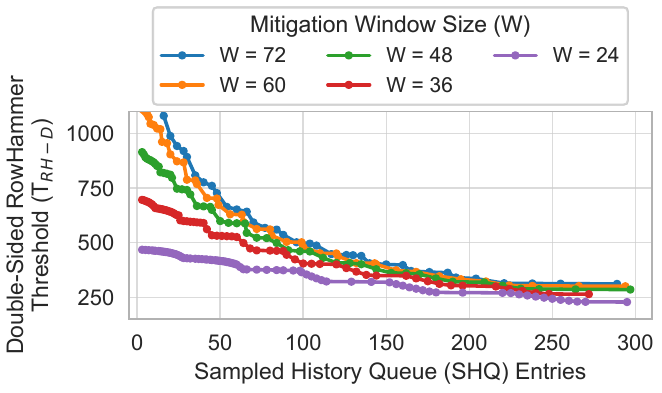}
    % \vspace{-0.14in}   
    \caption{Minimum Supported \TRHD{} of \DEFENSE{} under the worst-case circular-$X$-rows attack with varying SHQ size and $W$. Smaller $W$ lowers the supported \TRHD{} at low SHQ capacities, but its benefit diminishes once the SHQ provides enough history for repeated samples to intersect with high probability.}
    \label{fig:impact-w}
    \vspace{-0.15in}
\end{figure}

\topic{Impact of $W$}
\cref{fig:impact-w} shows the supported \TRHD{} as we vary the Sampled History Queue (SHQ) size and window size ($W$), with $R$ swept subject to $W\!\geq\!4R$ (\cref{subsec:prism-jedec}). Smaller $W$ lowers the supported \TRHD{} at low SHQ capacities by raising both the default mitigation rate ($1/W$) and the probability that an aggressor is sampled into the SHQ. For example, at $\sim\!40$ SHQ entries, $W\!=\!24$ supports \TRHD{} of 425, while $W\!=\!72$ remains above 700. However, this benefit diminishes at larger SHQ sizes ($\geq\!150$ entries), where the SHQ already provides enough history for persistent aggressors to intersect with high probability. Smaller $W$ is therefore most useful for ultra-low \TRHD{} under a constrained SHQ budget.

\smallskip\topic{Impact of \PMQTH{}}
The analysis so far has assumed that a row selected for mitigation is mitigated immediately, yielding the base threshold \TRHDBASE{}. In \DEFENSE{}, however, a selected row may remain in the Pending Mitigation Queue (PMQ) until it is serviced by a default mitigation (TRR or RFM) or through Alert Back-Off (ABO) protocol. While the row resides in the PMQ, it can accumulate further activations up to the \emph{tardiness threshold} (\PMQTH{}) before \ALERT{} is asserted. Thus, the PMQ-aware supported threshold becomes \TRHD{} = \TRHDBASE{} + \PMQTH{}. We use \PMQTH{}$=4$ as the default.

\smallskip\topic{Impact of activations during Alert Back-Off (\ABOACT{})}
\DEFENSE{} uses the existing JEDEC ABO protocol with one \RFM{} per ABO. The protocol allows up to 3 activations (\ABOACT{}) before the corresponding \RFM{} is issued. Moreover, a subsequent \ALERT{} can only be raised once the controller receives the activations that match the number of \RFM{}s during ABO. With a PMQ size of $Q$, an attacker can exploit these slack activations by preparing $Q\!-\!1$ entries near \PMQTH{}, then forcing a sequence of chained \ALERT{}s, analogous to the \emph{Feinting}~\cite{ProTRR} and \emph{Wave}~\cite{wave} attacks against PRAC~\cite{qprac, qureshi2024moat}. This chain accumulates up to \ABOACTQ{} additional activations on a targeted row before it is mitigated. The supported threshold of \DEFENSE{} therefore becomes \TRHD{} = \TRHDBASE{} + \PMQTH{} + \ABOACTQ{}. For our default PMQ size of 16, \ABOACTQ{} is 12, which combined with \PMQTH{} of 4 gives \TRHD{} = \TRHDBASE{} + 16.  

\subsection{Determining PrISM Configurations}\label{subsec:prism-config}
\DEFENSE{} can support a target \TRHD{} through different combinations of $(W, R, L)$. These parameters trade off performance and storage in different ways. Increasing $R$ sharply lowers \TRHD{} by raising both per-window sampling and the probability of intersections, allowing \DEFENSE{} to support lower \TRHD{} with fewer SHQ entries. However, larger $R$ also increases the maximum number of possible intersections per window ($R\!-\!1$), which can raise \ALERT{} frequency and degrade performance. It can also make \DEFENSE{} more vulnerable to potential \emph{Denial-of-Service (DoS)} attacks, in which an adversary intentionally causes frequent intersections to trigger repeated \ALERT{}s (see~\cref{sec:prism-dos} for a detailed DoS analysis). Smaller $W$ similarly reduces the required SHQ storage, particularly when the SHQ budget is tight, as shown in~\cref{fig:impact-w}. However, a smaller $W$ also leads to more frequent proactive \RFM{}s, causing higher slowdown. We therefore choose the default configuration that balances performance and storage.

\cref{table:prism-config} shows the selected \DEFENSE{} configurations. We use $W\!=\!72$ whenever the target threshold can be met without an overly large $R$, preserving the lowest proactive \RFM{} rate for \TRHD{} $\geq$ 500. For ultra-low \TRHD{} of 250, we reduce $W$ to 48 to keep the SHQ size practical without relying on a large $R$. Overall, \DEFENSE{} maintains practical hardware overhead. For \TRHD{} of 1000, it requires only a 36-entry SHQ, comparable to the TRR tracker already deployed in commercial DRAM~\cite{jattke2021blacksmith}. Even at a \TRHD{} of 250, \DEFENSE{} requires 632 entries, which remain 10--100$\times$ smaller than prior secure in-DRAM mitigations such as Mithril~\cite{kim2022mithril} and ProTRR~\cite{ProTRR} at the same threshold.

\begin{table}[h!]
\centering
\caption{\DEFENSE{} Configurations for Target RowHammer Thresholds}
\begin{footnotesize}
\resizebox{\columnwidth}{!}{%
\begin{tabular}{cccccc}
\toprule
\textbf{\begin{tabular}[c]{@{}c@{}}Target\\\TRHD{}\end{tabular}} &
\textbf{\begin{tabular}[c]{@{}c@{}}Supported\\\TRHD{}\end{tabular}} &
\textbf{\begin{tabular}[c]{@{}c@{}}Window\\ Size ($W$)\end{tabular}} &
\textbf{\begin{tabular}[c]{@{}c@{}}Sampled\\ Slots ($R$)\end{tabular}} &
\textbf{\begin{tabular}[c]{@{}c@{}}Lookback\\ Window ($L$)\end{tabular}} &
\textbf{\begin{tabular}[c]{@{}c@{}}SHQ Entries\\ ($(R-1)\times L$)\end{tabular}} \\
\midrule
1000 & 975 & 72 & 4 & 12  & 36  \\
750  & 731 & 72 & 7 & 11  & 66  \\
500  & 499 & 72 & 7 & 41  & 246 \\
250  & 249 & 48 & 9 & 79 &  632 \\
\bottomrule
\end{tabular}
} % end resizebox
\end{footnotesize}
\label{table:prism-config}
\end{table}

\subsection{Sensitivity to Mean-Time-to-Failure (MTTF)}
We next evaluate how the supported \TRHD{} of our chosen \DEFENSE{} configurations changes under different MTTF targets. \cref{table:mttf_sens} reports the supported \TRHD{} for each configuration across a range of per-bank MTTF values, along with the corresponding system-level MTTF. We derive the system-level MTTF assuming an attacker targets up to $24$ banks in parallel out of our $32$-bank system, thereby maximizing the per-bank activation rate under the tFAW constraint. Targeting more banks would throttle the per-bank hammer rate and weaken the attack. \DEFENSE{} provides strong protection across a wide MTTF range with practical SHQ sizes. For example, raising the per-bank MTTF target from $10$K to $1$M years (system MTTF from $417$ to $41.7$K years) shifts the supported \TRHD{} of the $66$-entry configuration from $731$ to $786$, and the $36$-entry configuration from $975$ to $1069$, a small relaxation in the supported threshold for a $100\times$ improvement in MTTF.

\begin{table}[h!]
\centering
\caption{
Supported \TRHD{} of \DEFENSE{} with Varying Mean-Time-To-Failure
}
\begin{footnotesize}
\resizebox{\columnwidth}{!}{%
\begin{tabular}{cccccc}
\toprule
\multirow{3}{*}[-2ex]{\textbf{\shortstack{Target MTTF\\(Bank)}}} &
\multirow{3}{*}[-2ex]{\textbf{\shortstack{Target MTTF\\(System)}}} &
\multicolumn{4}{c}{\textbf{Selected \DEFENSE{} Configuration}} \\
\cmidrule(lr){3-6}
 & &
\textbf{\shortstack{1K cfg.\\36 SHQ}} &
\textbf{\shortstack{750 cfg.\\66 SHQ}} &
\textbf{\shortstack{500 cfg.\\246 SHQ}} &
\textbf{\shortstack{250 cfg.\\632 SHQ}} \\
\midrule
1K years   & 41.7 years  & 944  & 720 & 478 & 247 \\
\textbf{10K years}  & \textbf{417 years} & \textbf{975}  & \textbf{731} & \textbf{499} & \textbf{249} \\
100K years & 4.17K years & 1017 & 747 & 507 & 262 \\
1M years   & 41.7K years & 1069 & 786 & 525 & 274 \\
\bottomrule
\end{tabular}
}%
\end{footnotesize}
\label{table:mttf_sens}
\end{table}

\section{Evaluation Methodology}\label{sec:eval_methodology}
\topic{Simulation Framework}
We evaluate \DEFENSE{} using Ramulator2~\cite{kim2015ramulator,ramulator2}, a cycle-accurate, trace-based DRAM simulator. Following prior work~\cite{yauglikcci2021blockhammer, qprac, dapper, olgun2023abacus}, we employ Ramulator2's internal out-of-order (OoO) core model\footnote{We validated the internal core model by reproducing the PRAC interface-speed experiment in \cref{fig:motivation_interface} using the open-source ChampSim–Ramulator2 integration from prior work~\cite{prac_timing_channel_isca25, tprac_github}. Performance from the two setups agreed within 1.2\% across 3200–8000MT/s DDR5 speeds and showed matching trends, so we report results using Ramulator2's internal core model.}.
Our baseline system consists of an 8-core OoO processor, a 16MB shared LLC, and a single-channel, single-rank DDR5 configuration with 32GB of DRAM. The memory controller uses the Minimalist Open-Page address mapping~\cite{MOP} and a First-Ready First-Come First-Served scheduler~\cite{zuravleff1997controller, FRFCFS}. The DRAM is modeled as a 32Gb DDR5-8000B device with timing parameters from the JEDEC DDR5 specification~\cite{jedec_ddr5_prac}. \cref{table:system_config-ramulator2} shows the detailed system configuration.

\begin{figure*}[t]
    \centering
    \includegraphics[width=0.92\linewidth, height=\paperheight, keepaspectratio]{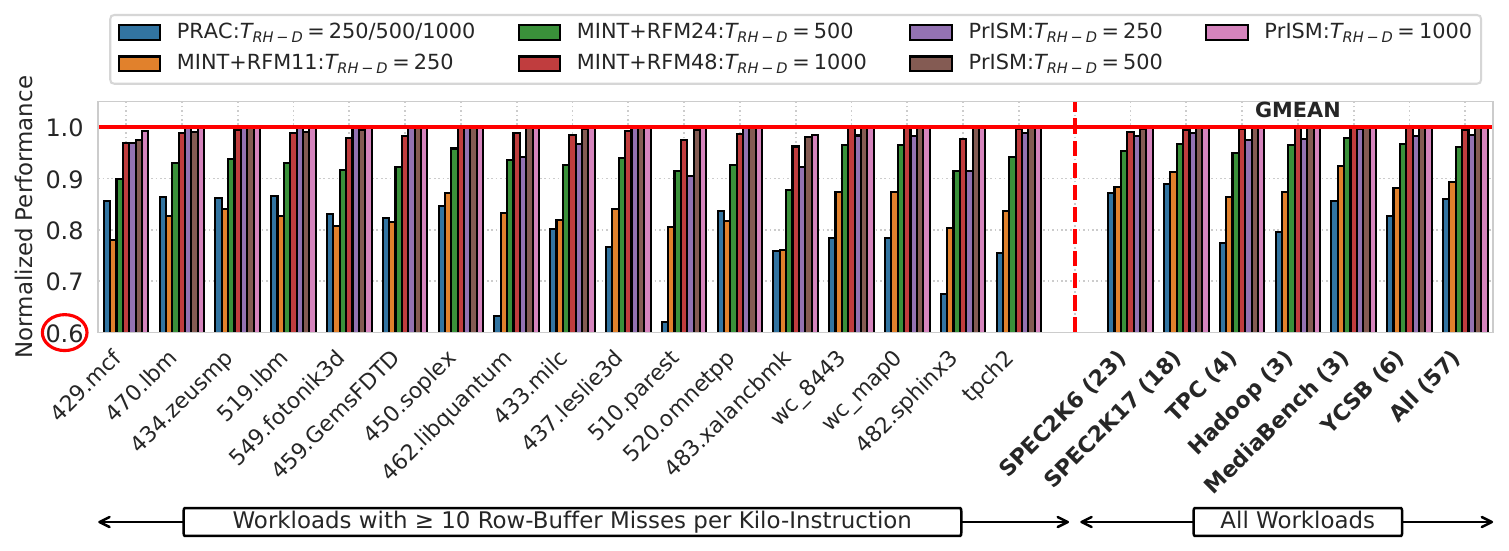}
    \vspace{-0.1in}
    \caption{Performance of \DEFENSE{} at \TRHD{} of 250, 500, and 1000, compared to MINT~\cite{MINT} and PRAC~\cite{qprac}. MINT incurs low slowdown at higher \TRHD{}, but its slowdown increases as \TRHD{} drops because lower thresholds require more frequent fixed-rate \RFM{}s to address the non-selection problem. PRAC incurs roughly 14\% average slowdown across thresholds because its overhead is dominated by inflated timing parameters. In contrast, \DEFENSE{} incurs negligible slowdown ($\leq\!0.2\%$) at \TRHD{}~$\geq 500$ and only 1.5\% at ultra-low \TRHD{} of 250, since intersections are infrequent on benign workloads.}
    \label{fig:perf_main}
\end{figure*}
\begin{figure*}
    \centering
    \vspace{-0.1in}
    \includegraphics[width=0.92\linewidth, height=\paperheight, keepaspectratio]{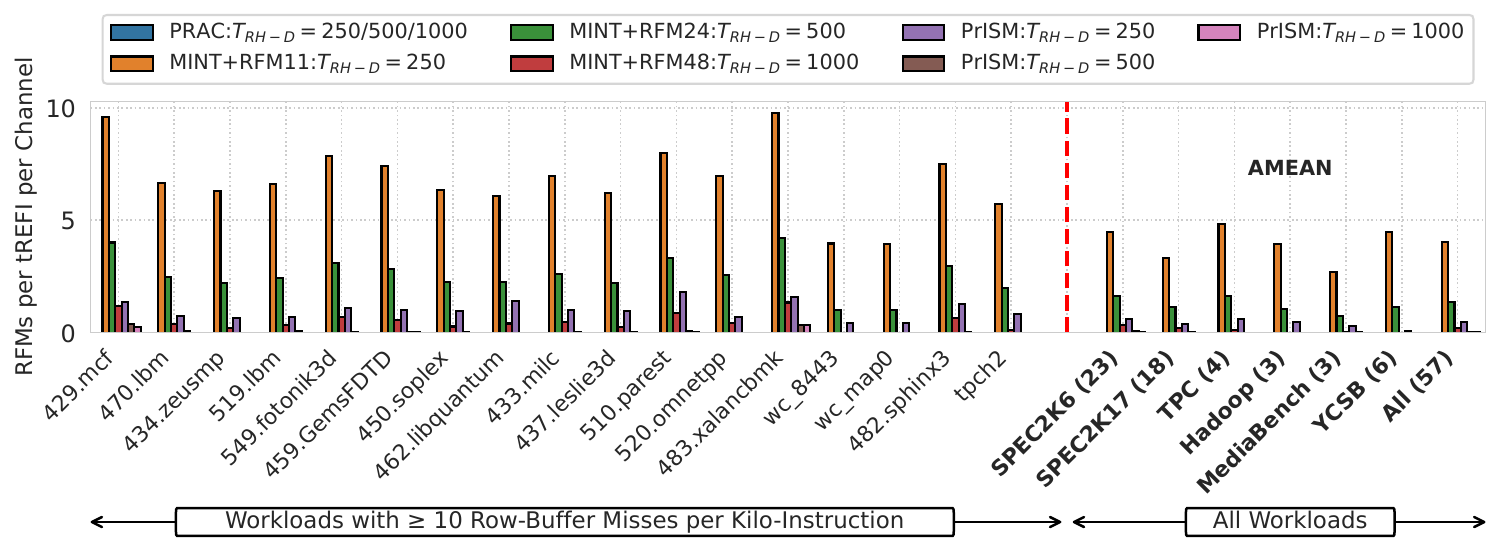}
    \vspace{-0.1in}
    \caption{RFM frequency per \TREFI{} per channel for \DEFENSE{}, MINT, and PRAC. MINT requires more frequent fixed-rate \RFM{}s as \TRHD{} drops, explaining its higher slowdown at low \TRHD{}. \DEFENSE{} issues fewer \RFM{}s by keeping the default mitigation rate low and requesting additional \RFM{}s through Alert Back-Off only on intersections. PRAC's precise per-row tracking yields negligible \RFM{}s but still incurs overhead from inflated timing parameters.}
    \label{fig:mitigation_rate}
    \vspace{-0.05in}
\end{figure*}

\begin{table}[t!]
\begin{center}
\begin{small}
\caption{System Configuration}{
\begin{footnotesize}
\resizebox{0.95\columnwidth}{!}{
\begin{tabular}{c|c}
\toprule
  Out-Of-Order Cores &  8 Cores, 4GHz, 4-wide, 512-entry ROB\\
  Last Level Cache (Shared)   & 16MB, 8-way, 32 MSHRs per core   \\\midrule
  Address Mapping & Minimalist Open-Page (MOP)~\cite{MOP}\\ 
  Scheduling Policy& FR-FCFS~\cite{zuravleff1997controller, FRFCFS} with a cap of 1~\cite{FRFCFS_CAP}\\\midrule
  Memory Type                  & 32Gb DDR5-8000B \\
  DRAM Organization      & 4 Bank $\times$ 8 Groups $\times$ 1 Rank $\times$ 1 Channel \\
  Rows Per Bank, Size                 & 128K, 8KB \\ 
  tRCD, tCL, tRAS & 16ns, 16ns, 32ns\\
  tRP, tRTP, tWR, \TRC{} & 16ns, 7.5ns, 30ns, 48ns \\ 
  \TRFC{}, \TREFI{}     &   410 ns, 3.9$\mu$s \\
  t\RFMAB{}, t\RFMSB{} & 350ns, 190ns \\
\bottomrule

\end{tabular}}
\end{footnotesize}
}
\label{table:system_config-ramulator2}
\end{small}
\end{center}
\end{table}
\begin{table}[t!]
\vspace{-0.12in}
\centering
\caption{Workload Categorization Based on RBMPKI}
% \vspace{-0.1in}
\label{table:workload_categorization}
\begin{footnotesize}
\resizebox{0.95\columnwidth}{!}{
\begin{tabular}{c|c}
\toprule
\textbf{RBMPKI} & \textbf{Workloads} \\ \midrule

% ================= HIGH =====================
\multirow{4}{*}{\shortstack{\textbf{High} \\ (17) \\ {[10+)}}}
& 429.mcf, 470.lbm, 434.zeusmp, 519.lbm, 549.fotonik3d, \\
& 459.GemsFDTD, 450.soplex, 462.libquantum, 433.milc, \\
& 437.leslie3d, 510.parest, 520.omnetpp, 483.xalancbmk, \\
& wc\_8443, wc\_map0, 482.sphinx3, tpch2 \\
\hline

% ================= MEDIUM =====================
\multirow{5}{*}{\shortstack{\textbf{Medium} \\ (21) \\ {[1, 10)}}}
& 471.omnetpp, grep\_map0, 473.astar, 505.mcf, tpch17, \\
& 436.cactusADM, jp2\_decode, 507.cactuBSSN, 557.xz, tpcc64, \\
& jp2\_encode, ycsb\_aserver, ycsb\_eserver, ycsb\_bserver, \\
& ycsb\_cserver, ycsb\_dserver, 500.perlbench, 523.xalancbmk, \\
& tpch6, ycsb\_abgsave, 456.hmmer \\
\hline

% ================= LOW =====================
\multirow{4}{*}{\shortstack{\textbf{Low} \\ (19) \\ {[0, 1)}}}
& 401.bzip2, 502.gcc, 435.gromacs, 458.sjeng, 445.gobmk, \\
& 525.x264, 508.namd, 531.deepsjeng, 544.nab, 526.blender, \\
& 403.gcc, 464.h264ref, h264\_encode, 447.dealII, 444.namd, \\
& 481.wrf, 541.leela, 538.imagick, 511.povray \\
\bottomrule
\end{tabular}
% \vspace{-0.2in}
}
\end{footnotesize}
\vspace{-0.1in}
\end{table}

\smallskip
\topic{Evaluated Design} 
We compare \DEFENSE{} against two in-DRAM RowHammer mitigations: (1) Per Row Activation Counting (PRAC)~\cite{jedec_ddr5_prac} and (2) the state-of-the-art probabilistic scheme MINT~\cite{MINT}.
For PRAC, we adopt QPRAC~\cite{qprac} as a secure baseline: each \ALERT{} triggers one \RFM{}, and a 5-entry priority service queue tracks aggressor candidates for mitigation. We tune the Back-Off threshold for each target \TRHD{} following QPRAC's publicly available security model.  
For MINT, we follow its design with a delayed mitigation queue to support refresh and \RFM{} postponement. 
The mitigation window size is set following the original work~\cite{MINT}: for \TRHD{} of $250$, $500$, and $1000$, the window is $11$, $24$, and $48$ activations, respectively. 

For \DEFENSE{}, we configure the mitigation window ($W$), the number of sampled slots ($R$), and lookback window ($L$) according to \cref{table:prism-config} for each target \TRHD{}. We use a 16-entry Pending Mitigation Queue (PMQ) by default.

\smallskip
\topic{Workloads} 
We evaluate 57 open-source workloads in Ramulator2~\cite{ramulator_opensource}, drawn from SPEC2006/2017~\cite{SPEC2006,SPEC2017}, TPC~\cite{TPC}, Hadoop~\cite{hadoop}, MediaBench~\cite{MediaBench}, and YCSB~\cite{ycsb}, classified by row-buffer misses per kilo-instruction (RBMPKI) as summarized in \cref{table:workload_categorization}.
We run 8-core homogeneous mixes, with each core executing 250 million instructions for 2 billion instructions per run. By default, we issue one Target Row Refresh (TRR) every two refresh intervals and use Same-Bank RFM (\RFMSB{}) for proactive \RFM{}. We also evaluate sensitivity to TRR frequency and PMQ size. Performance is reported as weighted speedup over a non-secure DDR5 baseline.

\section{Results and Analysis}
\subsection{Performance Overhead}
\cref{fig:perf_main} compares the performance of \DEFENSE{}, MINT, and PRAC across \TRHD{} values of 250, 500, and 1000. \DEFENSE{} achieves the lowest slowdown among the evaluated designs. For \TRHD{} $\!\geq\! 500$, \DEFENSE{} incurs negligible slowdown (under 1\%), and only 1.5\% even at the ultra-low \TRHD{} of 250. The modest increase at low \TRHD{} comes mainly from the shorter mitigation window, which increases the proactive \RFM{} rate. As shown in~\cref{fig:mitigation_rate}, \DEFENSE{} issues only 0.03 \RFM{}s per \TREFI{} at \TRHD{} of 500, and 0.45 at \TRHD{} of 250.

MINT has low overhead at higher thresholds, but its overhead grows as \TRHD{} drops because it raises its fixed mitigation rate to maintain security, reducing effective memory bandwidth. As~\cref{fig:mitigation_rate} shows, MINT issues only 0.2 \RFM{}s per \TREFI{} at \TRHD{} of 1000, but 4 at \TRHD{} of 250, leading to 3.8\% and 10.7\% average slowdown at \TRHD{} of 500 and 250, respectively. The impact is most pronounced on high-memory-intensity workloads. For example, on \emph{429.mcf}, MINT issues nearly 4 and 10 \RFM{}s per \TREFI{} at \TRHD{} of 500 and 250, causing about 10\% and 21.9\% slowdown, respectively, while \DEFENSE{} stays under 3.2\% on the same workload across all thresholds. At \TRHD{} of 250, MINT's slowdown even exceeds PRAC's on several workloads (\textit{e.g.,} \emph{470.lbm}, \emph{434.zeusmp}, \emph{520.omnetpp}). In contrast, \DEFENSE{}'s worst-case per-workload slowdown stays at 9.4\% (\emph{510.parest}) even at \TRHD{} of 250.

PRAC issues virtually no \RFM{}s thanks to its precise per-row tracking. However, its inflated timing parameters (\TRP{} and \TRC{}) from counter updates on every activation cause roughly 14\% average slowdown across thresholds, exceeding $20\%$ on several workloads such as \emph{510.parest} and \emph{wc\_8443}. Overall, \DEFENSE{} avoids the frequent fixed-rate \RFM{}s required by MINT at low \TRHD{} and the timing overhead of PRAC, providing the lowest slowdown among the evaluated designs, especially for high-memory-intensity workloads at low \TRHD{}.

\subsection{Sensitivity to Target Row Refresh Ratio}\label{subsec:results-senstrr}
\begin{figure}[h!]
\centering
\vspace{-0.05in}
\includegraphics[width=0.9\linewidth,height=\paperheight,keepaspectratio]{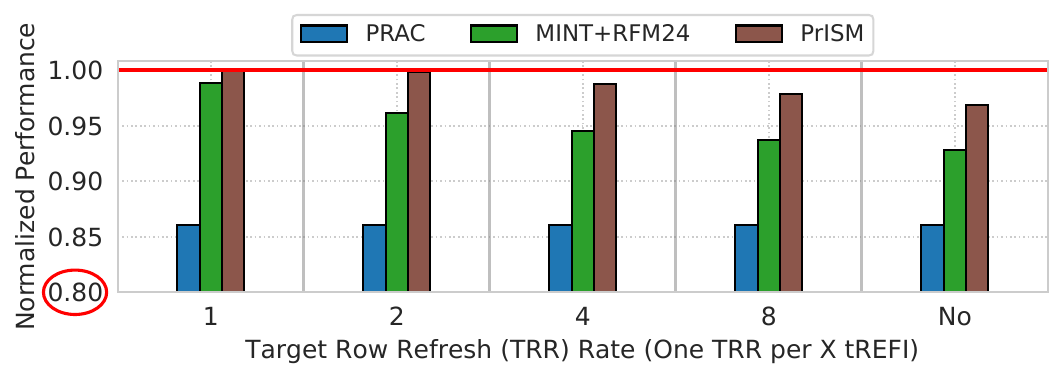}
\vspace{-0.1in}
\caption{Performance impact of TRR rate for \DEFENSE{}, PRAC, and MINT at \TRHD{} of 500. Lower TRR rates increase reliance on proactive RFMs, degrading MINT and \DEFENSE{} performance, while PRAC remains nearly constant because its slowdown stems primarily from increased timing parameters. Overall, \DEFENSE{} achieves the lowest slowdown across all TRR settings.}
\label{fig:tref_sens}
\vspace{-0.1in}
\end{figure}

\cref{fig:tref_sens} compares the performance of \DEFENSE{}, MINT, and PRAC as the Target Row Refresh (TRR) rate varies at \TRHD{} of 500. Lower TRR rates degrade both MINT and \DEFENSE{} because they require more proactive RFMs rather than piggybacking mitigations onto TRR. With one TRR per \TREFI{}, \DEFENSE{} incurs negligible slowdown ($<$0.1\%). Without TRR, \DEFENSE{} issues an \RFM{} roughly every 72 activations, increasing slowdown to 3.2\%. Similarly, MINT's slowdown increases from 1.2\% with one TRR per \TREFI{} to 7.2\% without TRR. In contrast, PRAC remains near 14\% slowdown across TRR settings because its overhead is dominated by increased timing parameters rather than mitigation frequency. Overall, \DEFENSE{} achieves the lowest slowdown across all TRR rates.

\subsection{Sensitivity to Pending Mitigation Queue Size}
\cref{tab:pmq-sensitivity} evaluates the impact of \DEFENSE{}'s Pending Mitigation Queue (PMQ) size at \TRHD{} of 500. A larger PMQ improves performance because rows selected through intersections can wait longer for default TRR or RFM opportunities, reducing the number of \ALERT{}s. However, it also increases storage and the worst-case activations an adversary can accumulate via delayed chained-\ALERT{} behavior (\cref{subsec:trhd_analysis}). Since performance saturates at a 16-entry PMQ (matching the 32-entry case at 0.2\%), we use 16 entries by default.

\begin{table}[h!]
\centering
\caption{Impact of PMQ Size on $ABO_{ACT}(Q)$ and Performance}
\resizebox{0.7\columnwidth}{!}{%
\begin{tabular}{ccc}
\toprule
\textbf{PMQ Size} & \textbf{$\textbf{ABO}_{\textbf{ACT}}(\textbf{Q})$} & \textbf{Performance Overhead} \\
\midrule
4  & 7  & 0.6\% \\
8  & 10 & 0.3\% \\
\textbf{16} & \textbf{12} & \textbf{0.2\%} \\
32 & 14 & 0.2\% \\
\bottomrule
\end{tabular}
}
\label{tab:pmq-sensitivity}
\end{table}

\subsection{Storage and Power Overhead}
\topic{Storage}
\DEFENSE{} uses three per-bank structures: a Sampled History Queue (SHQ), a 13-entry Sampled Slot Queue (SSQ), and a 16-entry Pending Mitigation Queue (PMQ). Each SHQ and SSQ entry stores a 17-bit row address and a valid bit (18 bits total), and each PMQ entry additionally stores a 3-bit saturating activation counter. At \TRHD{} of 1000, \DEFENSE{} requires a 36-entry SHQ, resulting in 152B of SRAM per bank; at \TRHD{} of 500, it requires 625B per bank.

This storage is substantially smaller than secure counter-based in-DRAM mitigations across all evaluated thresholds. At \TRHD{} of 1000, Mithril~\cite{kim2022mithril} requires 1082 27-bit entries even when issuing an \RFM{} every 16 activations, the highest RFM rate defined in the DDR5 specification~\cite{jedec_ddr5_prac}; this is roughly \textbf{24$\times$} more storage than \DEFENSE{}. ProTRR~\cite{PROHIT} requires more than 16K entries per bank at the same threshold, or roughly \textbf{363$\times$} more storage than \DEFENSE{}. The gap narrows at lower thresholds but remains substantial: Mithril requires \textbf{20$\times$} (\TRHD{} of 500) and \textbf{13$\times$} (\TRHD{} of 250) more storage than \DEFENSE{}, while ProTRR requires \textbf{170$\times$} and \textbf{154$\times$}, respectively. Thus, \DEFENSE{} provides secure RowHammer protection with low slowdown while requiring one to two orders of magnitude less storage than secure counter-based in-DRAM mitigations.

\smallskip
\topic{Power}
\DEFENSE{} refreshes up to 4 victim rows per selected aggressor. Using the Micron power calculator~\cite{micron:calc}, we estimate that these refreshes would account for 2.8\% of DRAM power if issued every \TREFI{}. However, even at \TRHD{} of 250, \DEFENSE{} triggers \ALERT{}-induced mitigations only once every 20 \TREFI{} on average, keeping refresh-related power overhead at 0.14\%.

\DEFENSE{} also requires random sampling, which can be implemented using an in-DRAM pseudo-random number generator (PRNG) or true random number generator (TRNG), as in prior in-DRAM probabilistic schemes~\cite{DSAC,jaleel2024pride,MINT}. Following MINT~\cite{MINT}, we assume a 7-bit TRNG~\cite{rng1,rng2} that consumes 90~$\mu$W of static and 200~$\mu$W of dynamic power (290~$\mu$W total), three orders of magnitude lower than DRAM chip power. We also estimate the dynamic power of \DEFENSE{}'s SRAM structures using CACTI-7.0~\cite{CACTI}: 1.5~mW at \TRHD{} of 500 and 2.4~mW at \TRHD{} of 250, corresponding to 0.6\% and 1.0\% of the 245~mW DRAM chip power estimated by the Micron power calculator~\cite{micron:calc}. Overall, \DEFENSE{} incurs low power overhead. \DEFENSE{} can also leverage recent in-DRAM TRNG designs~\cite{olgun2025trng1,olgun2021trng2,kim2019trng3,tehranipoor2016trng4}, which generate high-entropy bits from stochastic DRAM phenomena such as reduced-latency activation failures. \DEFENSE{}'s per-activation logic fits within \TRC{} and incurs no DRAM timing overhead.

\section{Analyzing DoS Vulnerability of PrISM}\label{sec:prism-dos}
\DEFENSE{} uses the existing Alert Back-Off (ABO) protocol to request additional mitigations when repeated sampled activity creates SHQ intersections. Since ABO stalls an entire DRAM channel while \RFM{}s are serviced, frequent \ALERT{}s can reduce bandwidth available to co-running applications. \DEFENSE{} also performs default mitigations through periodic proactive \RFM{}s, which consume additional bandwidth. An adversary can therefore exploit both sources of mitigation traffic to mount a \emph{denial-of-service (DoS)} attack, similar to prior performance attacks on RowHammer defenses~\cite{dapper, roguerfm, rfm-sidechannel_usenix, breakhammer2024}. We now analyze \DEFENSE{}'s worst-case DoS exposure.

\subsection{Attack Model and Assumptions}
\DEFENSE{} has low overhead on benign workloads because SHQ intersections are infrequent, so additional mitigations are requested only occasionally. Instead, a DoS adversary tries to maximize the number of intersections. We model this adversary using the Circular-$X$-Rows pattern from~\cref{sec:security_analysis}: the attacker repeatedly activates $X$ rows in round-robin order so that sampled rows reappear in the SHQ and create frequent intersections. In the worst case, each mitigation window triggers one default \RFM{} and up to $R\!-\!1$ additional \RFM{}s from intersections, totaling at most $R$ \RFM{}s per $W$ activations.

We assume proactive \RFMAB{} for default mitigation and disable Target Row Refresh (TRR), which is the worst case for \DEFENSE{} because mitigations cannot be hidden within refresh time. Following prior work~\cite{mopac_isca25, qureshi2024moat, qprac}, we use activation throughput as the DoS metric. Let $C_{\mathrm{RFM}}$ denote the cost of one \RFMAB{} in activation-slot units:
\begin{equation}
    C_{\mathrm{RFM}} = \frac{t_{\mathrm{RFMab}}}{t_{\mathrm{RC}}}
\end{equation}
With $t_{\mathrm{RFMab}}\!=\!350$ns and $t_{\mathrm{RC}}\!=\!48$ns, one \RFMAB{} stalls roughly seven activations. Thus, if \DEFENSE{} issues at most $R$ \RFM{}s per $W$ activations, the worst-case throughput loss is:
\begin{equation}
    \text{Throughput Loss}
    =
    \frac{C_{\mathrm{RFM}} \cdot R}{W + C_{\mathrm{RFM}} \cdot R}
    \approx
    \frac{7R}{W + 7R}
\end{equation}
This bounds the worst-case throughput loss from default and intersection-induced mitigations under sustained attacks.

\subsection{Worst-Case Slowdown under Performance Attacks}\label{subsec:prism-dos-impact}
\cref{tab:prism-dos} reports the worst-case slowdown of the selected \DEFENSE{} configurations under the Circular-$X$ DoS pattern. The default configurations incur worst-case slowdown factors of 1.39$\times$, 1.68$\times$, and 2.31$\times$ for \TRHD{} targets of 1000, 500, and 250, respectively. These slowdowns are comparable to existing memory performance attacks, such as conventional row-buffer conflict attacks~\cite{FRFCFS_CAP, memory_perf_attack1}. Thus, exploiting \DEFENSE{} does not create a fundamentally more severe DoS vector than those already present in multi-core systems with shared memory.

\begin{table}[h!]
\centering
\vspace{-0.05in}
\caption{
Worst-case DoS Slowdown of Selected \DEFENSE{} Configurations
}
\begin{footnotesize}
\resizebox{0.75\columnwidth}{!}{%
\begin{tabular}{cccc}
\toprule
\textbf{Target \TRHD{}} &
\textbf{$W$} &
\textbf{$R$} &
\textbf{Worst-Case Slowdown} \\
\midrule
1000 & 72 & 4 & 1.39$\times$ \\
500  & 72 & 7 & 1.68$\times$ \\
250  & 48 & 9 & 2.31$\times$ \\
\bottomrule
\end{tabular}
}
\end{footnotesize}
\label{tab:prism-dos}
\end{table}

This bound also exposes a performance-storage trade-off. Reducing $R$ limits the maximum number of intersection-induced \RFM{}s per window, improving DoS robustness. However, a smaller $R$ also lowers the probability that persistent aggressors create SHQ intersections, so the system must increase the lookback window $L$ to maintain the same security guarantee, increasing SHQ storage. Systems that prioritize stricter Quality-of-Service (QoS) guarantees can therefore use smaller-$R$, larger-$L$ configurations, trading additional storage for lower worst-case slowdown.

\section{Related Work}\label{sec:related_work}
\subsection{Alert-Based RowHammer Mitigations}
Several prior works use \ALERT{} for RowHammer mitigations. TWiCe~\cite{lee2019twice} adds counters in the register clock driver and uses \ALERT{} to request mitigations from the memory controller, but incurs substantial area overhead even at a \TRHD{} of 5K. Self-Managing DRAM~\cite{hassan2024self} and AutoRFM~\cite{autorfm_hpca25} leverage internal subarray structures to perform mitigations inside DRAM and redefine \ALERT{} to block activations to subarrays under mitigation. These approaches require significant changes to the DRAM core, limiting commercial adoption. In contrast, \DEFENSE{} reuses the existing Alert Back-Off (ABO) protocol~\cite{jedec_ddr5_prac} and operates without DRAM array or interface modifications. 

Concurrent work, MIRZA~\cite{MIRZA}, also uses ABO to reduce the overhead of static-rate probabilistic mitigation. MIRZA uses per-subarray counters as a coarse-grained activity filter and enables probabilistic mitigation only when a subarray exceeds a predefined activation threshold. This reduces unnecessary mitigations, but its effectiveness depends on how rows are distributed across subarrays; MIRZA therefore proposes a strided row-to-subarray mapping to improve this distribution. In contrast, \DEFENSE{} does not depend on a specific row-to-subarray mapping, since it correlates sampled row addresses directly across mitigation windows.

\subsection{Comparison with Other PRAC Designs}
\begin{figure}[h!]
\vspace{-0.1in}
\centering
\includegraphics[width=0.9\linewidth,height=\paperheight,keepaspectratio]{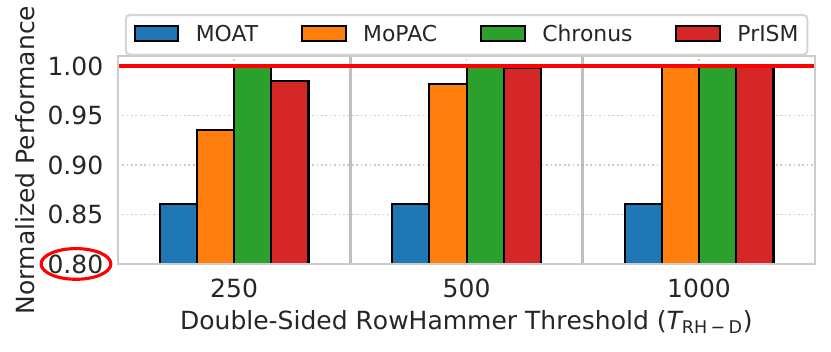}
\vspace{-0.13in}
\caption{Performance comparison of \DEFENSE{}, MOAT~\cite{qureshi2024moat}, MoPAC~\cite{mopac_isca25}, and Chronus~\cite{Chronus}. MOAT incurs 14\% slowdown from PRAC's per-activation counter updates, while MoPAC's slowdown grows at lower \TRHD{} due to more frequent sampled updates. \DEFENSE{} incurs negligible slowdown ($\leq\!0.2\%$) at \TRHD{} $\geq 500$ and only 1.5\% at \TRHD{} of 250, comparable to Chronus without requiring per-row counters or DRAM core changes.}
\vspace{-0.12in}
\label{fig:nrh_sens_compare}
\end{figure}

\cref{fig:nrh_sens_compare} compares \DEFENSE{} with recent secure PRAC-based mitigations: MOAT~\cite{qureshi2024moat}, Chronus~\cite{Chronus}, and MoPAC~\cite{mopac_isca25}. We configure each design following its paper for each target \TRHD{}. We assume one TRR opportunity per two \TREFI{} intervals, and use two drains per \TREFI{} for MoPAC.

The key difference is that all three prior designs remain fundamentally PRAC-based, relying on per-row activation counters. As a result, they inherit PRAC's area and power costs~\cite{jedec_ddr5_prac, DSAC}, while requiring mechanisms to either perform or hide counter updates. MOAT deterministically updates counters on every activation and therefore inherits PRAC's timing bottleneck, causing a 14\% average slowdown. Chronus avoids this bottleneck by updating counters concurrently with row accesses via a heterogeneous subarray architecture, resulting in negligible slowdown. However, Chronus requires DRAM core changes and can significantly tighten power constraints, potentially requiring a doubled tFAW budget~\cite{mopac_isca25}.

MoPAC reduces counter update overhead by probabilistically updating counters. It samples a subset of activations, buffers the selected rows in a queue, and drains the pending counter updates through TRR opportunities or \ALERT{}s. However, as \TRHD{} drops, MoPAC must sample more frequently: roughly every 8 activations at \TRHD{} of 500 and every 4 activations at \TRHD{} of 250. This increases Alert frequency because the update queue fills more often, and drained updates cause PRAC counters to reach the Alert threshold sooner. Consequently, MoPAC incurs 1.8\% average slowdown at \TRHD{} of 500 and 6.5\% at \TRHD{} of 250. In contrast, \DEFENSE{} incurs only 1.5\% average slowdown even at an ultra-low \TRHD{} of 250, while avoiding costly per-row activation counters entirely.

\subsection{Performance-Degradation Attacks and Mitigations}
Recent work has shown that RowHammer mitigations can be exploited for performance-degradation attacks~\cite{roguerfm, rfm-sidechannel_usenix}. DAPPER~\cite{dapper} employs secure hashing-based randomization, BreakHammer~\cite{breakhammer2024} throttles hardware threads that trigger excessive mitigation activity, and QPRAC~\cite{qprac} proposes to reduce interference from \ALERT{}s using finer-grained RFMs. Because \DEFENSE{} primarily focuses on designing a secure, scalable probabilistic mitigation, its operation is orthogonal to these dedicated DoS defenses. These techniques are complementary to \DEFENSE{}: access throttling can limit the triggers for adversarial mitigation, while finer-grained \RFM{}s can reduce \DEFENSE{}'s mitigation cost caused by \ALERT{}s.

\subsection{Host-Side RowHammer Mitigations}
Host-side mitigations rely on the memory controller to select rows for mitigation and issue targeted refreshes, such as Directed \RFM{}~\cite{jedec_ddr5_prac}. They either trigger mitigations probabilistically~\cite{kim2014architectural,kim2014flipping,PROHIT,MRLOC,hammerfilter,DREAM_isca25} or use activation trackers to identify aggressor rows~\cite{CBT,park2020graphene,qureshi2022hydra,comet,olgun2023abacus,saxena2024start}. Probabilistic schemes must increase mitigation frequency as RowHammer thresholds decrease, while tracker-based schemes require substantial storage to track many aggressors. In addition, host-side schemes may require additional support for transitive attacks, such as Half-Double~\cite{HalfDouble}, since the controller lacks proprietary internal DRAM mapping information. In contrast, \DEFENSE{} performs mitigation selection inside DRAM using sampled history, enabling scalable protection at low thresholds without host-side row tracking.

\subsection{Alternative Mitigation Mechanisms}
Row-migration techniques~\cite{saileshwar2022RRS,AQUA,CROW,ShadowHPCA23}, such as SRS~\cite{SRS} and AQUA~\cite{AQUA}, mitigate RowHammer by relocating frequently activated rows. Access-throttling mechanisms~\cite{breakhammer2024, yauglikcci2021blockhammer}, such as BlockHammer~\cite{yauglikcci2021blockhammer}, instead limit the activation rate of aggressive rows. However, these approaches can incur significant area or performance overhead as \TRHD{} drops below 1000. In contrast, \DEFENSE{} preserves low slowdown at low thresholds by using in-DRAM sampled history to request additional mitigations only for repeated sampled activity.

\subsection{ECC-Based RowHammer Defenses}
ECC-based defenses~\cite{ali2022safeguard,twobirds,RowArmor} can tolerate some RowHammer-induced bit flips by correcting corrupted data after errors occur. However, modern DRAM devices can exhibit multiple bit flips within a single ECC codeword~\cite{yauglikcci2024spatial,kim2020revisitingRH,jattke2021blacksmith}, exceeding the correction capability of conventional ECC. Prior work has also demonstrated successful RowHammer attacks on ECC-protected DRAMs~\cite{ecc_fail,cojocar2019eccploit}, including recent DDR5 devices~\cite{meyer2026phoenix}. In contrast, \DEFENSE{} prevents RowHammer by mitigating aggressor rows before faults occur, providing protection even at ultra-low \TRHD{}.

\subsection{DRAM Architecture and Interface Redesigns}
Other approaches propose fundamental redesigns of the DRAM architecture and memory interface to allow mitigations or refreshes to execute concurrently with DRAM activations~\cite{REGA_SP23, HIRA}. However, these schemes require modifications to the DRAM core and interface, hindering their commercial adoption. In contrast, \DEFENSE{} achieves low-\TRHD{} protection by reusing the existing ABO protocol~\cite{jedec_ddr5_prac}, avoiding DRAM array and interface changes.

\subsection{Timing Channel Attacks on RowHammer Defenses}
Recent work has shown that PRAC and RFM-based mitigations can introduce timing side channels because their \RFM{} timing can depend on memory activity~\cite{prac_tc_micro,prac_timing_channel_isca25,rfm-sidechannel_usenix}. TPRAC~\cite{prac_timing_channel_isca25} addresses this problem by making externally visible \RFM{} timing independent of aggressor activity. Timing-channel protection is orthogonal to our goal, and \DEFENSE{} can be composed with TPRAC-style techniques by decoupling mitigation selection from externally visible mitigation timing: intersections can still determine which rows enter the PMQ, while RFMs are released at fixed, activity-independent times.
\section{Conclusion}
We presented PrISM, a scalable in-DRAM probabilistic mitigation that resolves the non-selection problem in prior probabilistic RowHammer defenses. \DEFENSE{} correlates sampled rows across mitigation windows using a Sampled History Queue (SHQ), requesting additional mitigations only when repeated sampled activity creates intersections. This increases mitigation for persistent aggressors without uniformly increasing the default \RFM{} rate. \DEFENSE{} is compatible with the existing JEDEC Alert Back-Off protocol, requires no DRAM array changes, and avoids PRAC's per-row activation counters that must be updated on every activation. Our evaluations show that \DEFENSE{} securely supports \TRHD{} of 500 with a negligible 0.2\% average slowdown and 625B of SRAM per bank, and scales to an ultra-low \TRHD{} of 250 with 1.5\% average slowdown. Overall, \DEFENSE{} improves low-threshold scalability over fixed-rate probabilistic defenses while providing lower performance overhead than PRAC and one-to-two orders of magnitude less storage than secure counter-based in-DRAM mitigations.

\section*{Acknowledgment}
We thank the anonymous ISCA 2026 reviewers for their valuable comments. We are especially grateful to Moinuddin Qureshi and Reviewer-F for their detailed feedback, which significantly strengthened the final version of this paper. We also thank the Advanced Research Computing (ARC) team at the University of British Columbia (UBC) for their computing support~\cite{sockeye}. This work was supported by the Natural Sciences and Engineering Research Council of Canada (NSERC) under funding number RGPIN-2019-05059. The views expressed are those of the authors and do not necessarily reflect those of NSERC, NVIDIA, UBC, or the Government of Canada.

\begin{appendices}
\section{PRAC Performance Comparison}\label{app:prac-real-system}
Prior work~\cite{prac_perf_cal} evaluated PRAC on a real DDR5-4800 system and reported a 1.3\% slowdown on SPEC2017, whereas we observe an average 3.7\% slowdown at the same interface speed. This difference mainly stems from workload intensity and system configuration. First, our evaluated SPEC2017 workloads have higher RBMPKI, making them more sensitive to PRAC's increased timing parameters. Second, the prior study used 8 P-cores across two memory channels, corresponding to 4 cores per channel, whereas our setup uses 8 cores in a single channel. This 8-core-per-channel configuration is closer to modern server-class CPUs, which commonly provide roughly 8--16 physical cores per memory channel~\cite{amd_llc_size,intel_llc_size_288v}. Finally, the prior system provides 8.5MB of effective cache capacity per core, substantially reducing memory traffic and masking PRAC overhead. In contrast, we provision a server-representative 2MB of LLC per core~\cite{amd_llc_size,intel_llc_size_288v}. Recent studies~\cite{chrouns_arxiv,MIRZA} make similar observations.

\section{Analytical Model of SHQ Occupancy}\label{app:detailed_security}
This appendix derives the steady-state SHQ-presence probability used in~\cref{sec:security_analysis}. The model captures whether a repeatedly activated row is present in the Sampled History Queue (SHQ), which determines whether a future sampled activation creates an intersection. We use a two-state discrete-time Markov chain: $S_0$ denotes that the row is absent from the SHQ, and $S_1$ denotes that the row is present.

A transition from $S_0$ to $S_1$ occurs when the row is sampled in a mitigation window but is not selected as the default mitigation candidate, so it is inserted into the SHQ. For a row that appears once in a $W$-activation window, \DEFENSE{} samples the row with probability $R/W$. Given that it is sampled, one of the $R$ sampled rows is selected for default mitigation, so the row is inserted into the SHQ with probability $(R-1)/R$. Thus, the effective per-window insertion probability is:
\begin{equation}
P_{\text{in}} = \frac{R}{W}\cdot\frac{R-1}{R}
= \frac{R-1}{W}
\end{equation}
A transition from $S_1$ to $S_0$ occurs when the row's SHQ entry expires. Since each entry persists for $L$ lookback windows, we approximate the per-window exit probability as $1/L$.

At steady state, the insertion and eviction rates are equal:
\begin{equation*}
P(S_0)\cdot P_{\text{in}} = P(S_1)\cdot \frac{1}{L}.
\end{equation*}
Using $P(S_0)=1-P(S_1)$ and denoting $P(S_1)=P_{\text{SHQ}}$, we obtain:
\begin{equation}
P_{\text{SHQ}} =
\frac{L\cdot P_{\text{in}}}{1+L\cdot P_{\text{in}}}.
\end{equation}
This model gives the steady-state SHQ residency probability for a repeated aggressor. We use this value as $P_{\text{SHQ}}$ in~\cref{sec:security_analysis} to compute intersection-based mitigation probability.

\section{Extending PrISM to RowPress and ColumnDisturb}\label{app:rowpress_columndisturb}
\DEFENSE{} can be extended to support other read-disturbance mechanisms such as RowPress~\cite{rowpress} and ColumnDisturb~\cite{columndisturb_micro25}.

\smallskip
\topic{RowPress~\cite{rowpress}} 
\DEFENSE{} can mitigate RowPress by adopting the ImPress~\cite{impress} technique, as also used by MINT~\cite{MINT}. Each row-open interval is converted into an equivalent activation count, $EACT=(t_{ON}+t_{PRE})/t_{RC}$. \DEFENSE{} then advances its per-window activation counter by $EACT$ instead of one. If this increment crosses a sampled slot, the open row is explicitly sampled. This allows \DEFENSE{} to capture time-based aggressors using the same SHQ-intersection logic.

\smallskip
\topic{ColumnDisturb~\cite{columndisturb_micro25}} 
ColumnDisturb can disturb rows across the aggressor subarray and neighboring subarrays, so mitigation requires refreshing many potential victim rows rather than only adjacent rows~\cite{columndisturb_micro25}. \DEFENSE{} can serve as an efficient filter for identifying repeatedly activated aggressors and triggering mitigation mechanisms such as Proactively Refreshing Victim Rows (PRVR)~\cite{columndisturb_micro25}.
\end{appendices}
%%%%%%% -- PAPER CONTENT ENDS -- %%%%%%%%

%%%%%%%%% -- BIB STYLE AND FILE -- %%%%%%%%
\bibliographystyle{IEEEtranS}
\balance
\bibliography{main}
%%%%%%%%%%%%%%%%%%%%%%%%%%%%%%%%%%%%
\end{document}